\documentclass[conference]{IEEEtran}
\usepackage{cite}
\usepackage{amsmath,amssymb,amsfonts}
\usepackage{algorithmic}
\usepackage{graphicx}
\usepackage{textcomp}
\usepackage{booktabs}
\usepackage{siunitx}
\usepackage{svg}
\usepackage{float}

\usepackage{url} 

\def\BibTeX{{\rm B\kern-.05em{\sc i\kern-.025em b}\kern-.08em
    T\kern-.1667em\lower.7ex\hbox{E}\kern-.125emX}}

             \usepackage{fancyhdr}   

\fancypagestyle{firstpage}{     
  \fancyhf{}
  \chead{\textcolor{gray}{This article has been accepted for publication in the \\ 2025 IEEE Sensors Applications Symposium.}}
  \fancyfoot[C]{\small{\textcolor{gray}{~\copyright~ 2025 IEEE.  Personal use of this material is permitted.  Permission from IEEE must be obtained for all other uses, in any current or future media, including reprinting/republishing this material for advertising or promotional purposes, creating new collective works, for resale or redistribution to servers or lists, or reuse of any copyrighted component of this work in other works.}}}
}

\begin{document}

\title{Self-Sustaining Multi-Sensor LoRa-Based Activity Monitoring for Community Workout Parks}

\author{\IEEEauthorblockN{1\textsuperscript{st} Victor Luder}
\IEEEauthorblockA{\textit{PBL-DITET} \\
\textit{ETH Zurich}\\
Zurich, Switzerland \\
victor.luder@pbl.ee.ethz.ch}
\and
\IEEEauthorblockN{2\textsuperscript{nd} Michele Magno}
\IEEEauthorblockA{\textit{PBL-DITET} \\
\textit{ETH Zurich}\\
Zurich, Switzerland \\
michele.magno@pbl.ee.ethz.ch}
}

\maketitle

\begin{abstract}
With the rise of the Internet of Things (IoT), more sensors are deployed around us, covering a wide range of applications from industry and agriculture to urban environments such as smart cities. Throughout these applications the sensors collect data of various characteristics and support city planners and decision-makers in their work processes, ultimately maximizing the impact of public funds. This paper introduces the design and implementation of a self-sustaining wireless sensor node designed to continuously monitor the utilization of community street workout parks. The proposed sensor node monitors activity by leveraging acceleration data capturing micro-vibrations that propagate through the steel structures of the workout equipment. This allows us to detect activity duration with an average measured error of only 2.8 seconds. The sensor is optimized with an energy-aware, adaptive sampling and transmission algorithm which, in combination with the Long Range Wide Area Network (LoRaWAN), reduces power consumption to just $\mathbf{1.147\mskip3mu}$mW in normal operation and as low as $\mathbf{0.712\mskip3mu}$mW in low-power, standby mode allowing 46 days of battery runtime. In addition, the integrated energy-harvesting circuit was tested in the field. By monitoring the battery voltage for multiple days, it was shown that the sensor is capable of operating sustainably year-round without external power sources. To evaluate the sensor’s effectiveness, we conducted a week-long field test in Zurich, placing sensors at various street workout parks throughout the city. Analysis of the collected data revealed clear patterns in park usage depending on day and location, underscoring the value of this information for city planning and its potential social impact. This dataset is made publicly available through our online dashboard. Finally, we showcase the potential of IoT for city applications in combination with an accessible data interface for decision-makers.
\end{abstract}


\section{Introduction}
\thispagestyle{firstpage} 
The popularity of the Internet of Things (IoT) applications continues to grow, becoming increasingly important in our daily lives. This data spans a wide range of characteristics, such as localization~\cite{gnas_2024}, presence detection~\cite{zeng_2022}, and environmental sensing~\cite{ullo_2020}. IoT systems are applied across many domains, which include consumer products~\cite{luder_2024}, smart farming~\cite{Ali_2023, guzman_2022}, asset tracking in industrial contexts~\cite{priyanta_2019}, optimization of power distribution~\cite{Larrinaga_2021, metalliodu_2020}, and finally also for smart cities~\cite{kirimatat_2020, yang_2021, shen_2022}. The primary objective of such data collection is to provide end-users with data-driven insights that support informed decision-making. In both industrial and urban planning applications, these insights may help planners quantify the impact of their decisions.

With the trend of an ever-increasing number of sensors, the focus lies to reducing both maintenance efforts and operational costs. For a successful and sustainable deployment, it is essential to move towards compact, low-power, self-sustaining sensors~\cite{adu_2018, ma_2020}. Furthermore, these sensors should be integrated into existing wireless networks, uploading data to a database that offers easy access and preprocessing (e.g. filtering)  capabilities. Low Power Wide Area Networks (LPWAN) have been identified as particularly suitable for smart city deployments~\cite{noreen_2017}, because they are perfect for low-power data transmission. Nevertheless, energy efficiency is crucial for IoT sensor design, particularly in public spaces where frequent maintenance or battery replacement is not feasible~\cite{atul_2023}. By utilizing energy harvesting techniques, these sensors can operate autonomously, reducing their dependence on external power sources ~\cite{giordano_2023}. This approach in combination with the low-power requirements and wide coverage of the Long Range Wide Area Network (LoRaWAN), ensures long-term operation. 

This paper presents a low-power, self-sustaining IoT sensor designed to monitor activity anonymously at community street workout parks. These parks are publicly available and rely on taxpayer money. Through this paper, we show how the gathering of such data may support officials to justify public funding decisions. In collaboration with the city of Zurich and EWZ (Elektrizitätswerk der Stadt Zurich), provider of a city-wide LoRaWAN network, this paper conducts a real-world application field test, evaluating the sensor’s application in an urban setting.\newline

\begin{table*}[!t]
\centering
\caption{Overview LoRa based Monitoring System}
\label{relatedwork}
\begin{tabular}{@{}llll@{}} 
 \toprule
\textbf{Authors/ Year} & \textbf{Application} &  \textbf{Deployment} & \textbf{Sensors} \\
 \midrule
~\cite{schulthess_2023_1}-2023 & Cattle monitoring & Outdoor cow field for 3 days (2 sensors) & IMU, GNSS \\
 \midrule
~\cite{Codeluppi_2020}-2020 & Agriculture monitoring & 4 sensors on a farm (3 months) & humidity, temperature\\
 \midrule
~\cite{raja_2024}-2024 & Irrigation system & 10 sensors field test & temperature, humidity, soil moisture \\
 \midrule
~\cite{mayer_2023}-2023 & Indoor positioning system & Indoor \qty{200}{\square\meter} area for 700 days & IMU and UWB localization  \\
 \midrule
~\cite{Parri_2020}-2020 & Offshore sea farm monitoring & Open sea (\qty{8.33}{\kilo\meter}) & LoRa coverage test \\
 \midrule
~\cite{schulthess_2023_2}-2023 & Outdoor presence monitoring & 16 sensors for 2 months & IMU, microphone, GNSS, temperature\\
 \midrule
\textbf{Our application}  & \textbf{Workout park monitoring} & \textbf{Outdoor 7 days, city wide (3 sensors)} & \textbf{accelerometer, time-of-flight sensor}\\
\bottomrule
\end{tabular}
\vspace{-0.2cm}
\end{table*}

The main contributions of this work include:

\begin{itemize}
\item The design and implementation of an end-to-end LoRa-based sensing system to monitor street workout parks in an urban environment. This implementation incorporates custom sensor hardware and firmware along with a web-based user interface for easy data access.

\item Reduced energy consumption through energy-aware firmware, incorporating adaptive sampling and transmission, combined with electronics that are capable of solar energy harvesting, resulting in a year-round, self-sustaining operation and reduced maintenance requirements after deployment.

\item Successful deployment of a real-world field test at multiple street workout parks, using the city-wide LoRa network. 

\item Evaluation of a detailed energy consumption analysis that demonstrates and quantifies the self-sustaining capabilities of the device under real-world operating conditions.
\end{itemize}

\section{Related Work}
Table \ref{relatedwork} briefly enumerates a selection of previous endeavors on the topic of LoRa-based smart city data collection and monitoring applications. The presented overview provides insights into the deployment as well as the incorporated sensors. For instance, Codeluppi et al.~\cite{Codeluppi_2020} demonstrate an example application in the field of agricultural monitoring, including a final test deployment spanning three months. Their solution does not include customized electronics to achieve low-power consumption and, ultimately, self-sustaining capabilities. However, the mentioned paper illustrates the challenges of a real-world application, exposing the sensors to environmental factors, such as weather and limited LoRa network coverage. A similar application is provided by Gopal and colleagues~\cite{raja_2024} who introduce a wireless, data-driven irrigation system. Furthermore, the limitations of the LoRa network coverage in a real-world application test is further investigated by the paper published by Parri~\cite{Parri_2020}, which conducts coverage tests over extended distances using offshore sea farming as an example application. Nevertheless, Schulthess et al.~\cite{schulthess_2023_1} introduce an energy-aware sensor with custom electronics to monitor the movements and activities of cattle, providing valuable insights into power requirements.
Meyer and colleagues~\cite{mayer_2023} demonstrate how solar harvesting may support low-power sensor nodes for monitoring people, while Shulthess et al.~\cite{schulthess_2023_2} introduce a smart city use-case, monitoring chairs in public places. Nevertheless, Shizhen~\cite{sizhen_2023} showed how a reliable sports monitoring setup can be implemented in a fitness center.

The novelty presented in this paper is the design and testing of a robust, low-power IoT sensor for street workout park monitoring. In combination with an application scenario, power evaluation and a final field test, we show its functionality, autonomous working capabilities and highlight the potential social impact of such data collection. Through its final deployment, this paper stands out by showing how data driven city planning may help city planners and benefit communities, who depend on publicly funded infrastructure.

\section{System Overview}
This section provides a hardware-software system overview of the proposed solution. It outlines two primary aspects: The design and integration of the hardware for the low-power sensor, the energy-aware firmware implementation and signal processing. 

\begin{figure}[!t]
\begin{minipage}[t]{1.0\linewidth}
\centering
\includegraphics[width=0.68\textwidth]{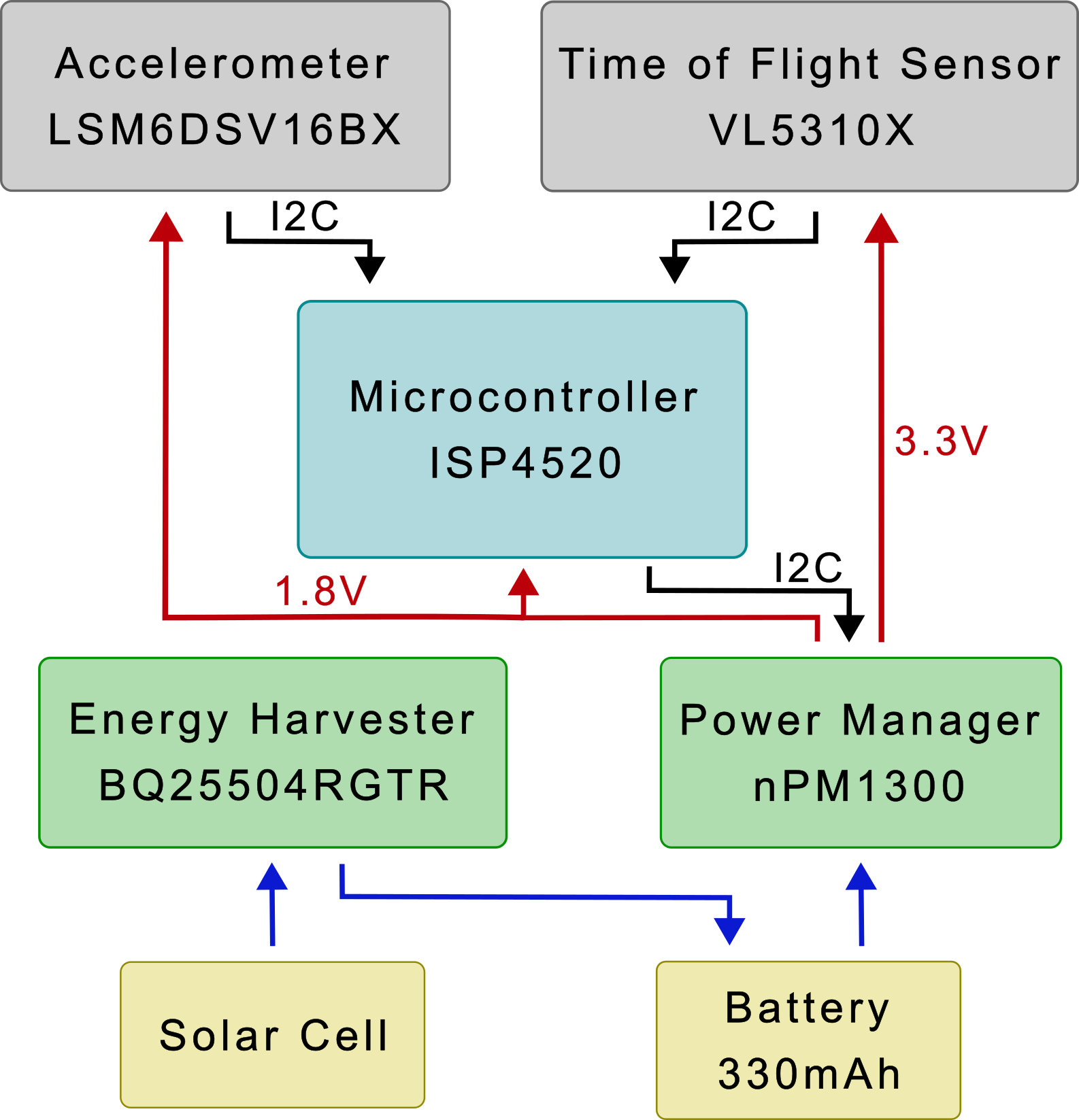}
\vspace{-0.1cm}
\caption{Hardware overview of the individual subsystems. Communication connections are marked in black, voltage supply in red and the power inputs from solar cells and battery in blue.}
\label{hardware_overview}
\end{minipage}
\vspace{-0.1cm}
\end{figure}

\begin{figure}
\begin{minipage}[t]{1.0\linewidth}
\centering
\includegraphics[width=0.75\textwidth]{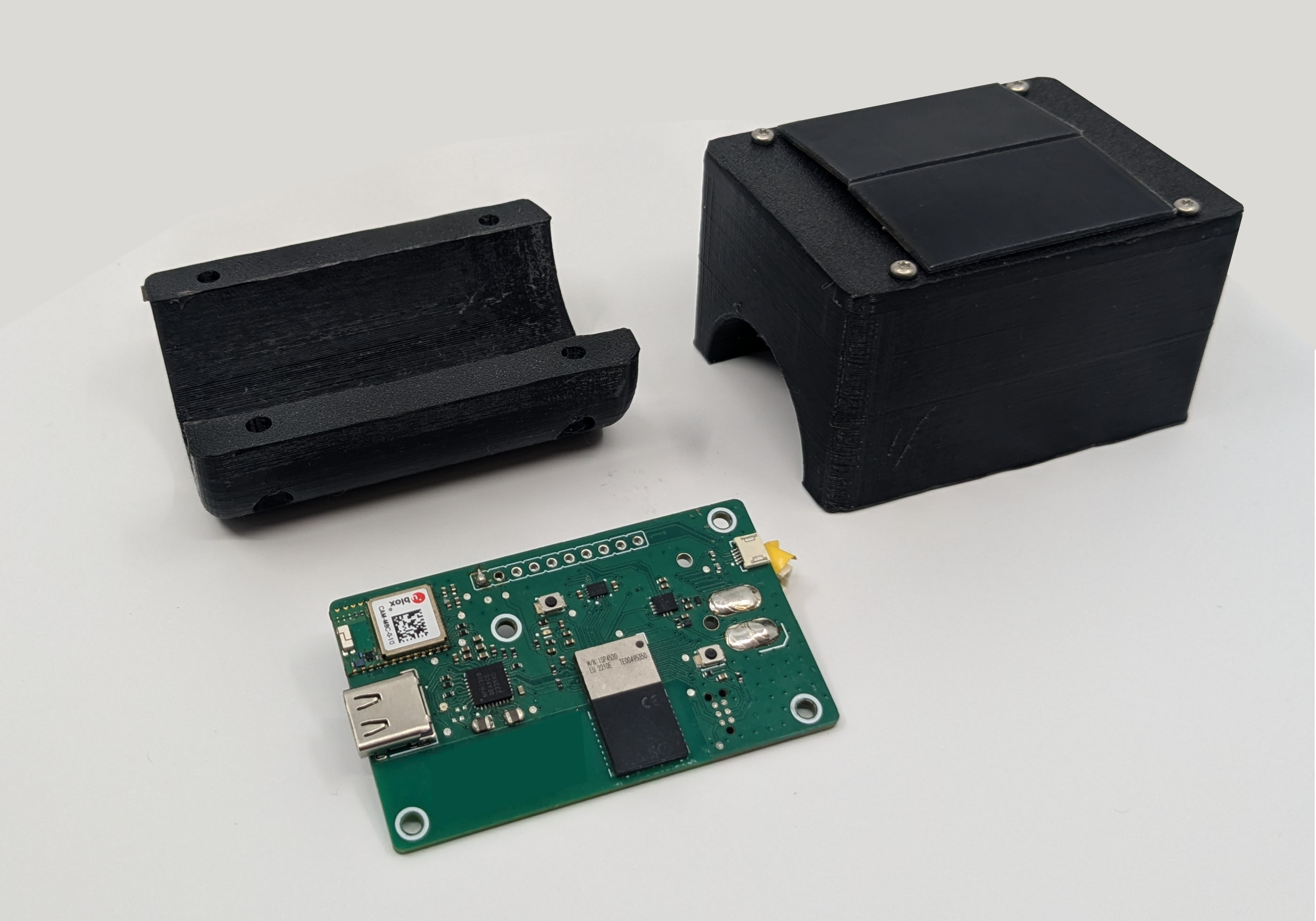}
\caption{The custom electronics along with the casing that fits the hand bar as well as the solar cells.}
\label{hardware}
\end{minipage}
\vspace{-0.3cm}
\end{figure}

\subsection{Hardware}
The sensing setup is built around the ISP4520 microcontroller from InsightSIP, which includes a low-power SX1261 Semtech LoRa module for efficient data transmission. To detect and measure activity, an accelerometer is used to pick up micro-vibrations propagating through the metal structure of the fitness equipment. For this purpose, the 3-axis Inertial Measurement Unit (IMU) from STMicroelectronics (LSM6DSV16BX), which features a high-performance mode for improved resolution, was selected.
To manage power efficiently, with a system voltage of \qty{3.3}{\volt} and \qty{1.8}{\volt}, the nPM1300 power manager, along with the BQ25504RGTR energy harvester, is chosen. This setup is capable of extracting energy from sunlight, with a minimum solar cell input voltage of just \qty{130}{\milli\volt}. For the solar cells, two high-efficiency KXOB201K04TF solar cells, with a 25\% efficiency operating at a MPP (Maximum Power Point) of \qty{2.23}{\volt}, are configured in parallel. 
Additionally, the VL5310X time-of-flight (ToF) sensor is used to add the feature of detecting the presence of individuals at a specific location, requiring an average of \qty{19}{\milli\ampere}~\cite{VL5310X_datasheet}. By power-gating the ToF sensor and activating it only when potential activity is detected, overall power consumption can be reduced.
To achieve a compact design, the system is powered by a \qty{330}{\milli\ampere{}\hour} lithium-ion battery. All components are integrated onto a custom PCB embedded within a casing specifically designed to fit the hand bars of the street workout park. The casing is optimized to allow perfect mechanical coupling with the structure, which is essential for detecting vibrations. Figures \ref{hardware_overview} and \ref{hardware} provide an overview of the hardware setup, while Figure \ref{deployed_sensor} displays the device when deployed in the field.

\begin{figure}[h!]
\vspace{-0.1cm}
\begin{minipage}[t]{1.0\linewidth}
\centering
\includegraphics[width=0.75\textwidth]{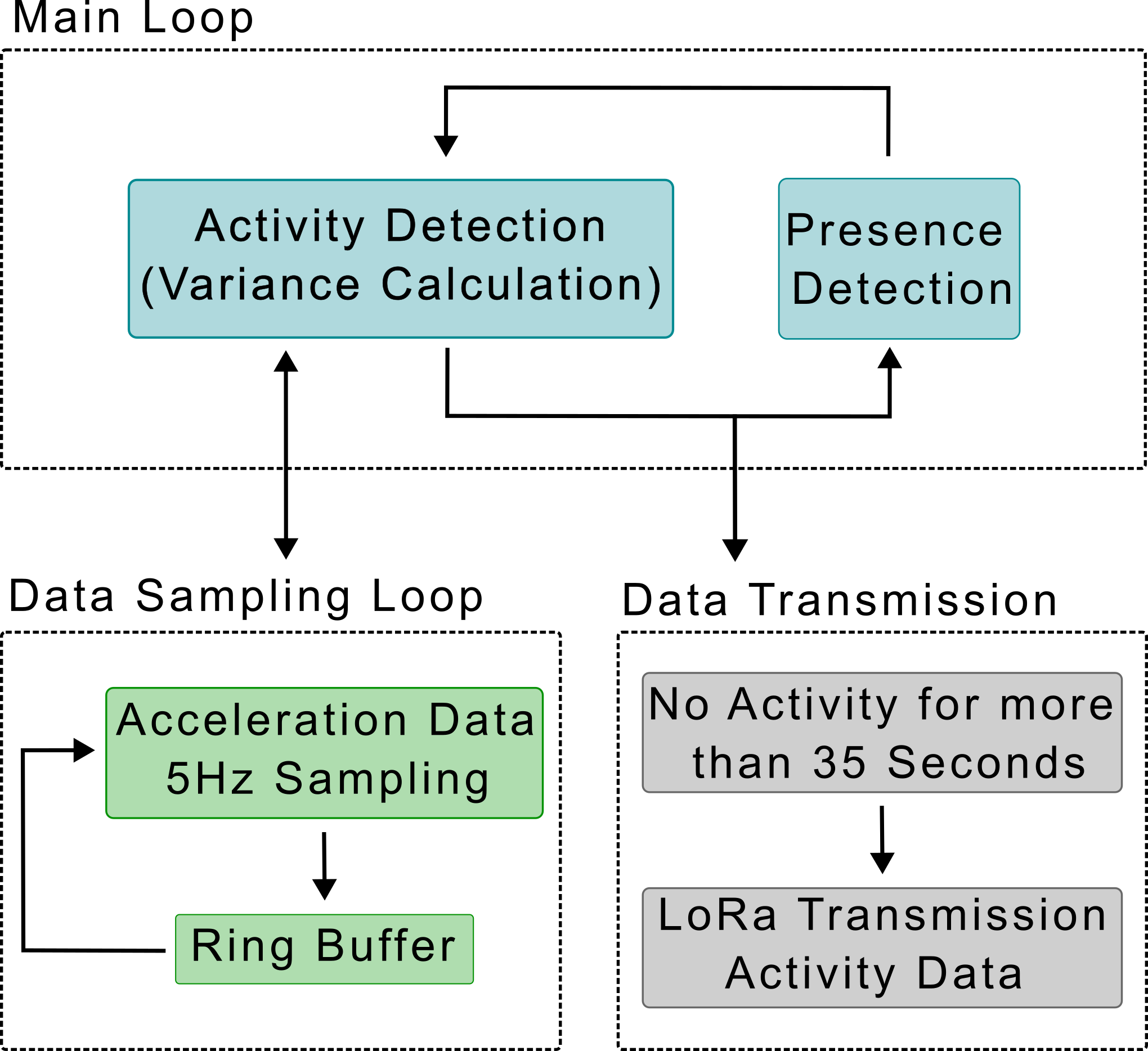}
\vspace{-0.1cm}
\caption{Firmware algorithm for reliable presence as well as activity detection.}
\label{software_overview}
\end{minipage}
\vspace{-0.35cm}
\end{figure}

\subsection{Activity Detection Algorithm}\label{activity_detection_section}
\begin{figure}[h!]
\begin{minipage}[t]{1.0\linewidth}
\centering
\includegraphics[width=1.0\textwidth]{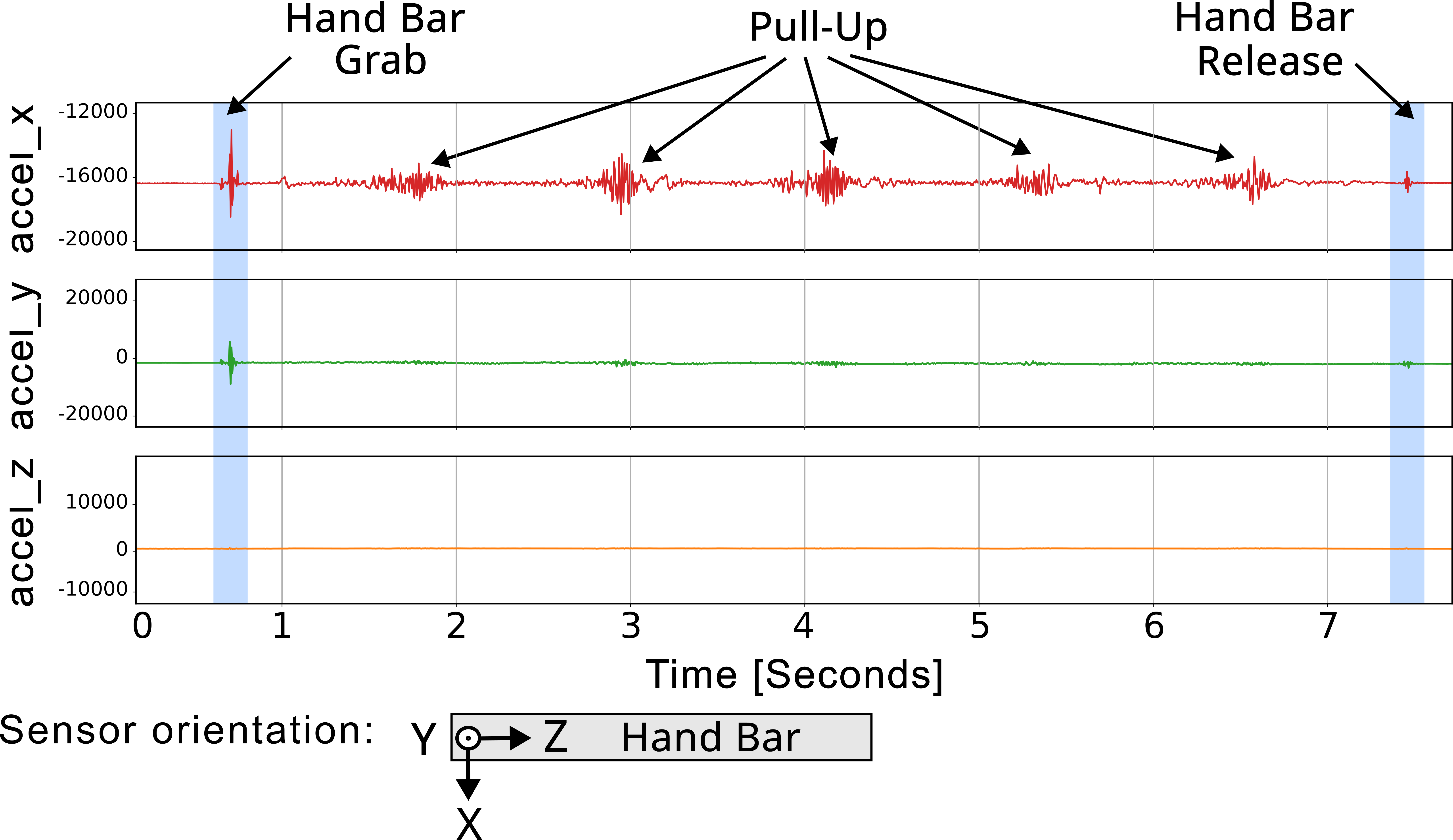}
\vspace{-0.6cm}
\caption{Acceleration data showing (raw values) the individual pull ups that can be clearly identified by an algorithm.}
\label{pull_ups}
\end{minipage}
\vspace{-0.3cm}
\end{figure}

To detect and monitor activity, the IMU is set to sample data at a low frequency of only \qty{5}{\hertz}, resulting in a total system power consumption of only \qty{0.712}{\milli\watt}. This frequency is chosen to allow the microcontroller to remain in sleep mode for as much of the time as possible, while the IMU is set to high-performance mode. The high-performance mode setting is essential since the higher noise level that might be picked up in low-power mode is critical for accurate detection. The sampled data is then continuously stored in a ten-sample ring buffer, which is used to determine the occurrence of an activity. 
Every second the algorithm evaluates the data in the ring buffer by calculating the variance across the sampled data. If the variance is elevated compared to intervals without detected activity, the algorithm marks the time instance as active. This approach leverages the oscillations of the fitness bars when touched, which is shown in Figure \ref{pull_ups}. It produces an acceleration signal similar to a spring oscillation, increasing the spread of the measured acceleration points. Given that a single pull-up takes at least one second, the \qty{5}{\hertz} sampling rate, in combination with the ten-sample buffer, allows the algorithm to reliably detect each pull-up with a \qty{1}{\hertz} resolution.
The algorithm is further designed to detect breaks in an exercise session. A break is registered if there is a calm period lasting at least 35 seconds, as repetitive exercising typically includes rest periods of no longer than 30 seconds. 
Once an activity with a total duration of at least 10 seconds is concluded, its duration is transmitted via LoRaWAN to a server, where it is stored in a database. This adaptive transmission strategy, sending data only when new data is available, reduces the number of energy-intensive data transmissions while ensuring no data is lost. Figure \ref{software_overview} provides a detailed illustration of the implementation. Additionally, the accelerometer’s sensitivity can be adjusted such that activities from any part of the workout park are registered. This is possible because the steel bars propagate vibrations through the structure and enables us to monitor overall park usage.

\subsection{Presence Detection Algorithm}
To complement the previously described algorithm, the acceleration data is combined with data from the VL5310X (time-of-flight) sensor, which is capable of detecting the presence of a person within a 2-meter range in a specified direction, reducing the risk of false positives. This sensor fusion approach enables the system to monitor overall activity at the fitness park while also pinpointing the presence of individuals at specific locations, preserving user anonymity throughout the process.
Furthermore, by cascading the detection into two stages, the algorithm remains in low-power operation. Initially, it remains in a standby state, monitoring for activity solely via the IMU as described in Section \ref{activity_detection_section}. Once an activity is detected, the algorithm activates the VL5310X sensor from sleep mode to check for the presence of a person, after which it returns the sensor to sleep mode again. As a result, each detected activity requires only approximately 0.08 seconds of active sensing to confirm the presence of an individual, significantly reducing the average power consumption for the presence detection mode to just \qty{6.951}{\milli\watt}.


\subsection{Software Infrastructure}
Given the increasing amount of data collected from numerous sensors distributed across the city, a user-friendly web interface is developed. This interface presents the data, that is stored in the database, to users with a visualization, along with filtering options for quick preprocessing (e.g. selection of sensor, selection of time span). 

 
\section{Experimental Evaluation}
To characterize the device in more detail, it must be determined whether it reliably and accurately detects activity. As an additional step, it is tested whether the time-of-flight sensor (VL5310X) detects the presence of a person at a given hand bar. Finally, the sensors must be tested in a field test to verify reliable data transmission and power management while enduring external real-world influences. Furthermore, the energy consumption of the sensor and its self-sustaining capabilities are evaluated and quantified.

\begin{figure}[]
\begin{minipage}[t]{1.0\linewidth}
\centering
\includegraphics[width=0.85\textwidth]{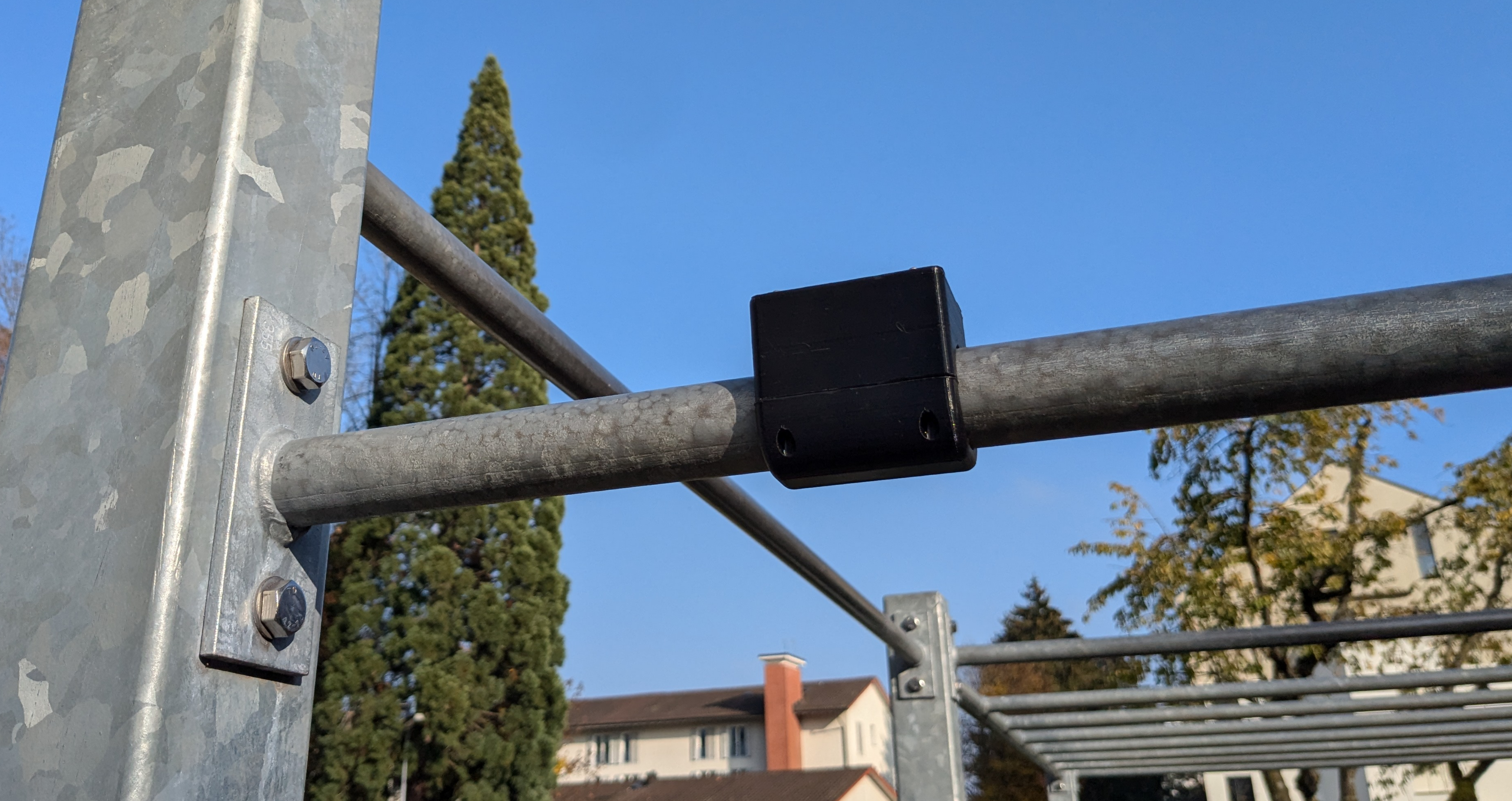}
\vspace{-0.2cm}
\caption{The final sensor in its casing deployed at a street workout park.}
\label{deployed_sensor}
\end{minipage}
\vspace{-0.5cm}
\end{figure}

\subsection{Activity and Presence Detection Evaluation}
To validate the accuracy of the setup, a sensor is installed on the pull-up bar at a street workout park in the city of Zurich, as shown in Figure \ref{deployed_sensor}. This placement allows the detection of activities when a person is exercising on the hand bar to which the sensor is attached, as well as any other activity performed in the park. To assess activity detection precision, ten exercise sessions were conducted, with activity duration recorded as a ground truth by observation (time-keeping) and via the sensor itself. Overall, the session durations ranged from 10 to 30 seconds, with an average difference of only 2.8 seconds. Additionally, a one-hour field test with regular park visitors was conducted to confirm accurate activity detection under real conditions, accounting for differences in individuals' exercise methods. The test demonstrated that activities involving multiple people across the park were detected with an average timing difference of 8.09 seconds. These activities lasted between 15 and 60 seconds and included rest breaks. Overall, this application test under real conditions showed that 100\% of the activities were detected and successfully transmitted via LoRaWAN to the web server, proving its reliability. 
Nevertheless, presence detection with the time-of-flight sensor (VL5310X) was also tested. For this purpose, 10 activities were conducted anywhere on the street workout park and 10 activities at a specific location where the person was expected to be detected. Overall, the sensor detected 100\% of the activities and correctly classified the activities whether they occurred at the hand bar where the sensor was installed or elsewhere in the park. The average timing error was at 4 seconds, while the activity durations ranged from 15 to 60 seconds. The combination of the time-of-flight sensor and the accelerometer may provide deeper insights into street workout park usage. 

\begin{figure}[!t]
\centering
\begin{minipage}[t]{1.0\linewidth}
\centering
\includegraphics[width=1\textwidth]{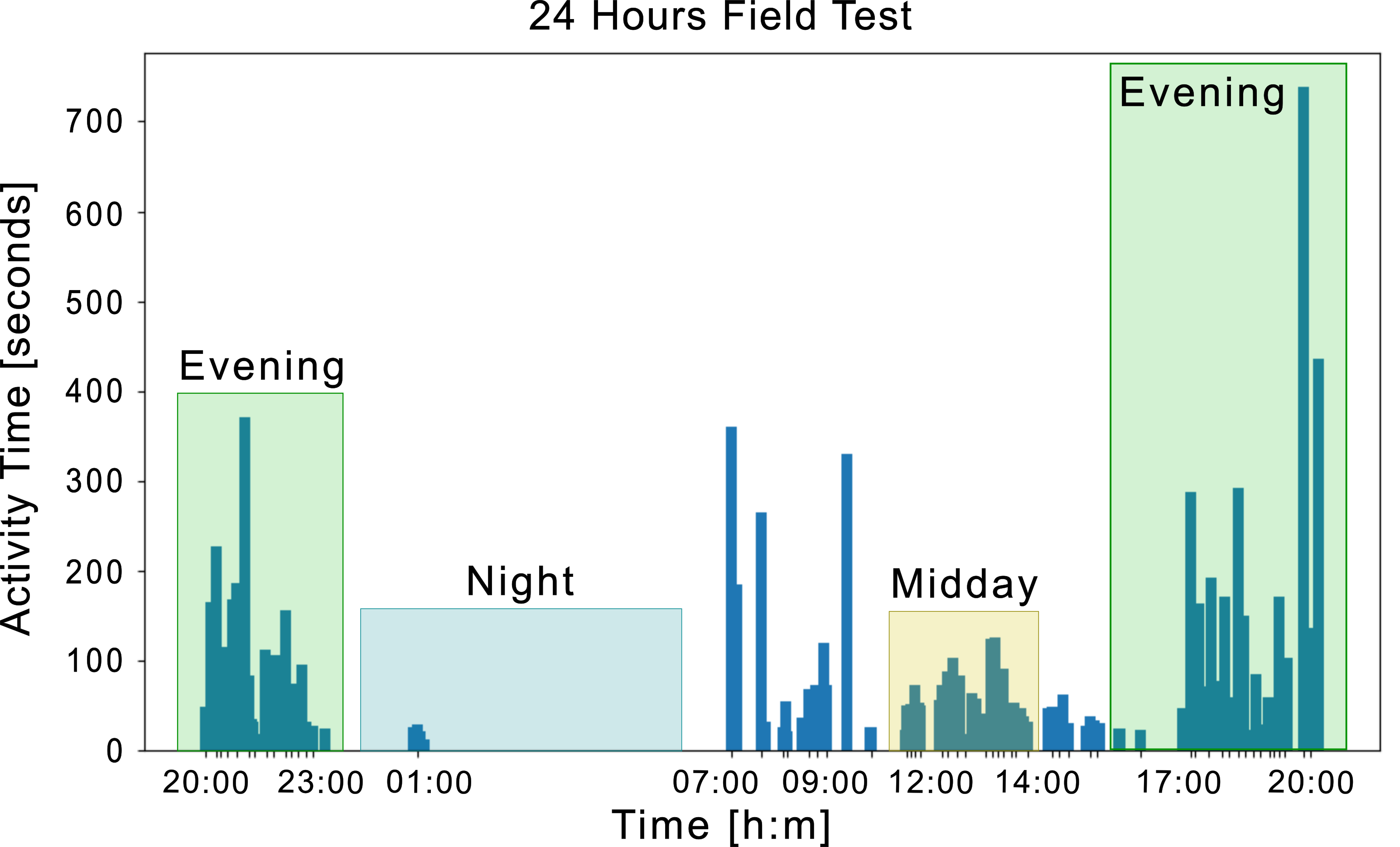}
\end{minipage}
\vspace{-0.8cm}
\caption{Activity time and occupancy of a street workout park over a total duration of 24 hours.}
\label{24_hours_field_test}
\vspace{-0.5cm}
\end{figure}

\subsection{Field Deployment (Application Test)}
\begin{figure}[!t]
\centering
\begin{minipage}[t]{1.0\linewidth}
\centering
\includegraphics[width=1\textwidth]{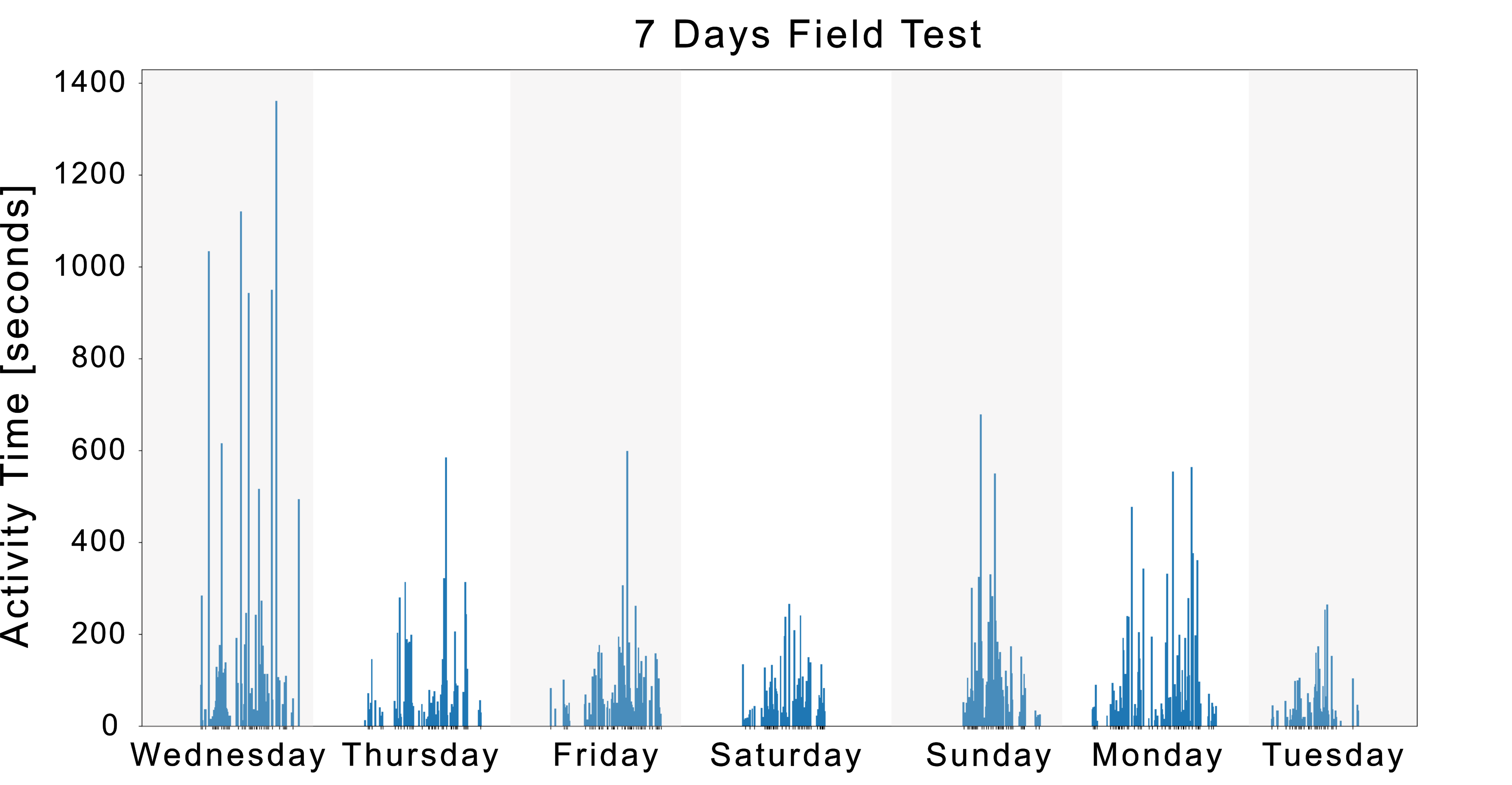}
\vspace{-0.8cm}
\end{minipage}
\caption{Activity data collected over the time of 7 days at one location.}
\label{one_week_field_test}
\end{figure}

For the conducted field experiments, sensors were installed at various street workout parks across the city. In collaboration with EWZ, which provides a city-wide LoRaWAN network, the sensors were integrated into the network, streaming the data in real time to a web server. In total, three sensors were deployed at three different locations. This enabled us to collect insights about usage patterns, such as peak times and occupancy throughout the day. Furthermore, we have a geographical resolution, which allows us to compare different areas within the city with varying demographic, financial and cultural characteristics. Figure \ref{24_hours_field_test} illustrates an example of a 24-hour activity profile for one of the parks. In Figure \ref{24_hours_field_test}, it can be seen that during the evening, the park is used the most. The very long activity time of up to 700 seconds suggests that multiple people are active simultaneously. Furthermore, the activities during midday are dense but short, indicating that a higher number of people utilizing the park, while the activity density during the morning and afternoon is reduced. 
Finally, to extend the application test, the sensors were deployed at multiple locations for a period of 7 days at each location. During this time, activity data was continuously collected and streamed to a server. The data provided valuable insights into the frequency and intensity of use at each street workout park. An example of this data for one park is shown in Figure \ref{one_week_field_test}.
The data reveals variation in activity depending on the day of the week, with lower usage on weekends and a noticeable shift in the earliest detected activity on Sundays. However, consistent patterns, such as peaks around midday and in the evening, can be observed across all days. Overall, the data shows that the park is actively used from early morning until late evening throughout the week, highlighting its value to the community.

\subsection{Power Consumption Evaluation}
To evaluate the power consumption and validate the sensor's self-sustaining capabilities, an analysis was conducted in two phases. First, the power consumption was measured in standby mode (with the accelerometer active and capable of detecting any activity). These measurements are complemented with measurements during data transmission, which represent the system's peak operational power. Using these values, the average expected power consumption was calculated, assuming an average battery voltage of \qty{3.9}{\volt}.

\begin{figure}[t]
\centering
\begin{minipage}[t]{1.0\linewidth}
\centering
\includegraphics[width=1\textwidth]{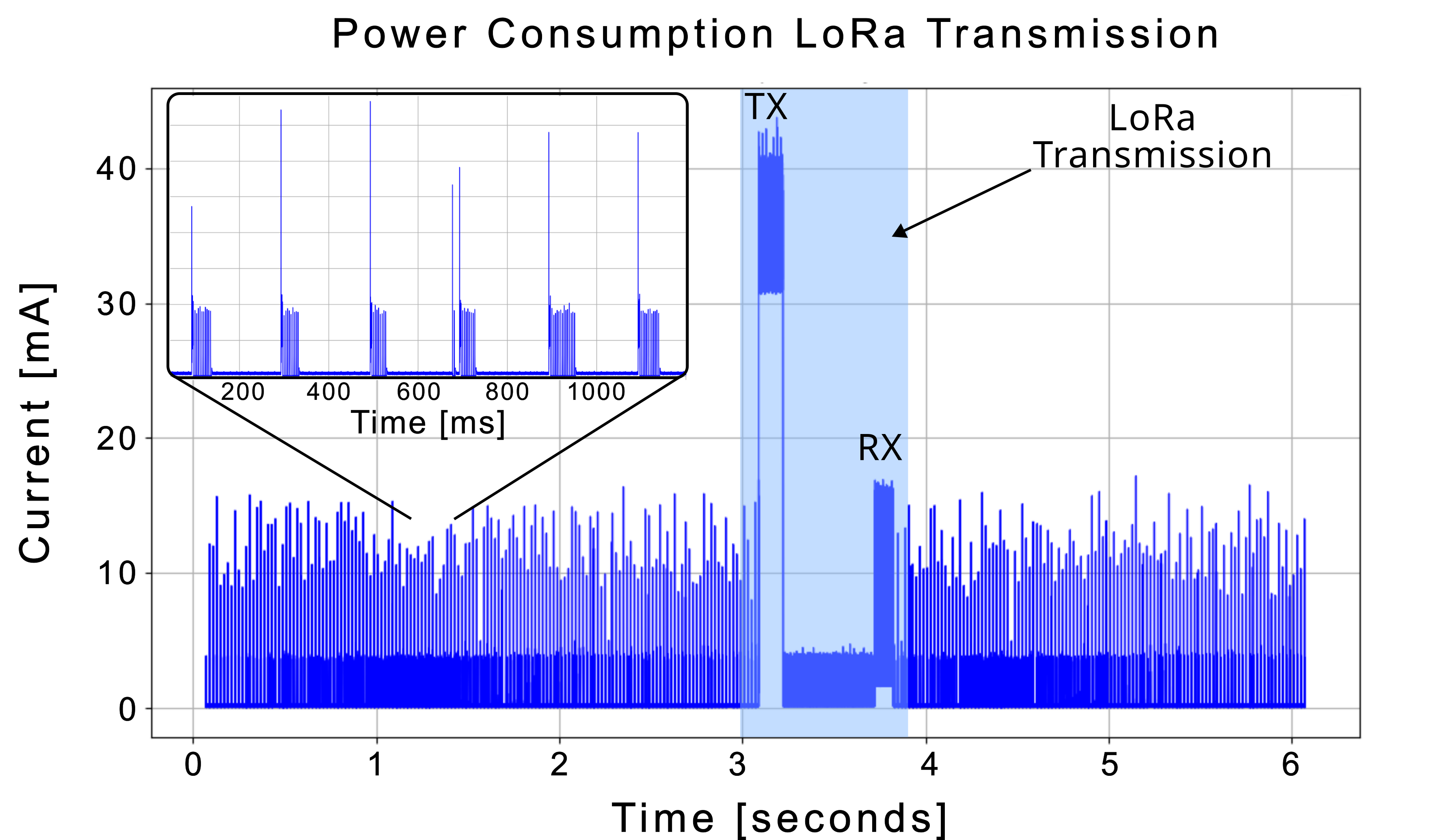}
\vspace{-0.6cm}
\end{minipage}
\caption{Power consumption when the system transmits data through LoRa. Looking closer into the measurement, it can be seen how the microcontroller repetitively returns to idle state reducing its current consumption to almost zero. }
\label{current_measurement}
\vspace{-0.5cm}
\end{figure}

\begin{figure}[h!]
\vspace{-0.1cm}
\centering
\begin{minipage}[t]{1.0\linewidth}
\centering
\includegraphics[width=0.98\textwidth]{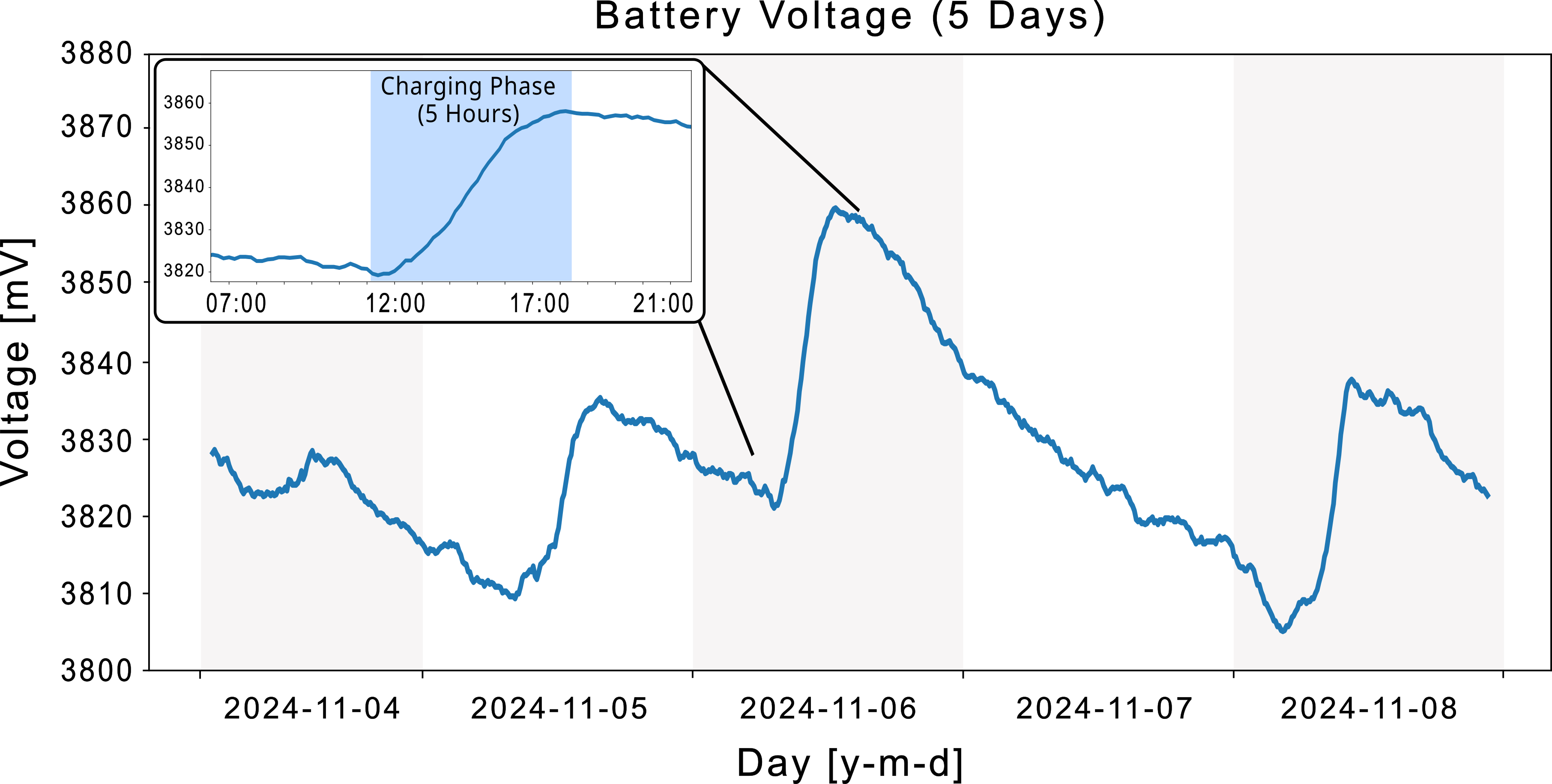}
\end{minipage}
\vspace{-0.5cm}
\caption{Voltage monitoring over a period of 5 days including days with no sun and days with sun.}
\label{voltage_battery}
\vspace{-0.3cm}
\end{figure}

In low-power mode, when no data is transmitted, the sensor has an average power consumption of \qty{0.712}{\milli\watt}. During the data transmission phase, with data sent every minute, average power consumption rises to \qty{4.194}{\milli\watt}. The measured current from this test is illustrated in Figure \ref{current_measurement}. Based on prior field tests, with approximately 180 activities expected per day, the average daily power consumption is calculated to be \qty{1.147}{\milli\watt}. Given a \qty{330}{\milli\ampere{}\hour} battery. This yields an estimated battery runtime of 46.75 days. A summary of the power measurements is provided in Table \ref{result_hardware}.
In a final evaluation, the battery voltage was monitored over a period of one week with the sensor connected to the energy harvester and two solar cells. Conducted during winter time, the evaluation faced limited sunlight and challenging external conditions. During the test, the acceleration sensor sampled as previously described at a rate of \qty{5}{\hertz}, while the LoRaWAN transmission period was set to 10 minutes. We evaluated the overall collected energy on a day with a sun exposure of 4.5 hours, which represents the average daily sun exposure in Zurich \cite{daylight_zurich}, resulting in an energy surplus of \qty{9.9}{\joule} during 24-hours of operation. This is further analyzed when evaluating how much energy surplus the system is capable to harvest in one hour of sun exposure. This results in a total of \qty[per-mode=symbol]{28.7}{\joule\per\hour} stored in the battery, while the system is fully operational and uploads data every 10 minutes. This measurement is illustrated in Figure \ref{voltage_battery}. It can be seen that on the 2024-11-04, as well as on the 2024-11-07, where little or no sun was present, the voltage decreased. Nevertheless, days with a total charging period of only 4.5 hours are enough to charge the battery, concluding the days with a positive overall balance. As verified with the energy balance, during the 5 days test, there was only a total of 13.5 hours of sun exposure (daily average of 2.7 hours), the test concluded the testing period on the same voltage level as it started. Overall, this test showed that even in challenging conditions, with less than average sun exposure the system may self-sustain itself, while the energy balance suggests that average sun exposure results in a positive daily energy balance, which allows year-round, self-sustaining operation without the need for recharging the battery.

\begin{table}[h]
\centering
\caption{Measured power consumption}
\label{result_hardware}
\resizebox{1.0\linewidth}{!}{
\begin{tabular}{@{}lr@{}} 
 \toprule
 & \textbf{Avg. Power}\\ 
 \midrule
 Data sampling mode  & \qty{0.712}{\milli\watt}  \\
 Data sampling and transmission  & \qty{4.194}{\milli\watt} \\ 
 Presence detection mode with VL5310X & \qty{6.951}{\milli\watt}\\
 Application mode & \qty{1.147}{\milli\watt}\\
 \midrule
Battery runtime (transmission every 10 minutes)   & 46.75 days\\
\bottomrule
\end{tabular}
}
\vspace{-0.4cm}
\end{table}

\section{Conclusion}
This paper introduces an energy-efficient, self-sustaining solution for the city-wide monitoring of street workout parks. Along with a web interface to access the data, the full system proves to be valuable to city decision-makers. With the integration of custom, low-power sensors into the city-wide LoRaWAN, along with energy harvesting capabilities and a user interface for data access, this paper covers all aspects of a low-maintenance, low-power smart city approach to efficiently gather and evaluate data. Furthermore, this setup has been tested in the field and proved to be robust. In the future, these sensors can be extended to collect environmental data, such as temperature and humidity, allowing the data analysis to be extended and placed into a wider context. Nevertheless, as the implementation of the time-of-flight sensor suggests, a more elaborated people-counting approach could be developed, adding even more value to the collected data. In the context of public street workout parks, this paper demonstrates how IoT sensors can be applied by decision-makers to measure the social impact of public funds, ultimately improving non-profit infrastructure, such as street workout parks, used by lower-income communities. 

\bibliographystyle{IEEEtran}
\bibliography{sample-base.bib}

\end{document}